\documentclass[12pt,preprint]{aastex}
\usepackage{natbib}
\usepackage{amssymb,amsmath} 

\shorttitle{GWs from Asymmetric GRBs}
\shortauthors{Du et al.}

\begin{document}

\title{Gravitational waves induced by the asymmetric jets of gamma-ray bursts}

\author{Shuang Du \altaffilmark{1},  Xiao-Dong Li \altaffilmark{1}, Yi-Ming Hu \altaffilmark{1}, Fang-Kun Peng \altaffilmark{2}, Miao Li \altaffilmark{1}}
 \altaffiltext{1}{Department of Physics and Astronomy, Sun Yat-Sen University, Zhuhai 519082, China; dushuang\_gxu@163.com, lixiaod25@mail.sysu.edu.cn}
 \altaffiltext{2}{Department of Physics And Electronics, Guizhou Normal University, Guiyang 550001, China}

\begin{abstract}
We study the gravitational wave (GW) production induced by the asymmetric jets of gamma-ray bursts (GRBs).
The asymmetric jets result in a recoil force acted on the central compact object,
whose motion leads to emission of GW.
Under reasonable assumptions and simplifications,
we derive the analytic form of the produce GWs.
The amplitude of emitted GWs is estimated to be relatively low,
but possibility exists that they can be detected by future experiments such as the Einstein Telescope.
We find the dynamical properties of the central object,
which is difficult to be studied via the electromagnetic (EW) channel,
can be inferred by measuring the emitted GWs.
Moreover, we find the emitted GWs can be used determine whether the relativistic jets is launched by
the neutrino annihilation process or the Blandford-Znajek process,
which cannot be clearly distinguished by the current GRB observations.
Our work manifests the importance of the GW channel in multi-messenger astronomy.
The physical information encoded in the GW and EW emissions of an astrophysical object is complementary to each other;
in case some physics can not be effectively investigated using the EW channel alone,
including the GW channel can be very helpful.
\end{abstract}


\section{Introduction}\label{sec1}

On 2017 August 17, the event GW170817 was detected by aLIGO,
and its electromagnetic counterparts were also detected by a serious of following experiments \citep{2017ApJ...848L..13A}.
The joint observation of gravitational waves (GWs) and electromagnetic waves (EWs) opens up the era of multi-messenger astronomy,
and could have significant impact on various research fields related with astronomy, gravity, cosmology,
high-energy physics, nuclear physics, and so on.

The information derived from the GWs and EWs (of the same source) can complement each other.
When using only one channel we can not obtain enough information to study some physical process,
including the other channel could be very helpful.
Especially, the GW channel is could be the only possible channel for solving some long-standing problems.
As an example, core-collapse supernovae (SNe) are optically opaque in the process of star collapse,
and only through GWs can we have an insight into the core area \citep{2016PhRvL.116o1102H}.

The situation is rather similar for the case of gamma-ray bursts (GRBs) physics.
GRBs originate from collapse events and merger events,
and can be associated with kilonovae and SNe \citep{1999ApJ...524..262M,2017Sci...358.1556C,2017Natur.551...64A,2017ApJ...848L..13A}.
The evolution of GRBs can be divided into three stages \citep{2015PhR...561....1K}:
1) compact binary merging or massive star collapsing
form the ``central engine'' (the system of a compact star surrounded by an accretion disk);
2) the central engines produce intermittent ultra-relativistic jets;
3) energy dissipation of the relativistic jets generates gamma-ray radiation.
The relativistic jets are believed to be launched by the neutrino annihilation process \citep{1999ApJ...518..356P,2017NewAR..79....1L}
or the Blandford-Znajek (BZ) process \citep{1977MNRAS.179..433B} of the central engines.
However, based on the electromagnetic observations,
it is difficult to tell which scenario is really happening in the GRBs.
Besides, the energy dissipation process of these jets is also an opening question \citep{1994ApJ...430L..93R,2011ApJ...726...90Z}.

It was proposed that the hint for the type of central engine can be obtained through the observation of GW ring-down \citep{2017ApJ...851L..16A}.
If the central compact object is found to be a neutron star (NS),
the relativistic jets are very likely launched by the neutrino annihilation process.
The reason is that, NSs have no ergospheres and thus are incompatible with the BZ mechanism\footnote{Some authors also point out that the NS with a stiff equation of state,
which has an ergosphere, can power jets via the BZ mechanism \citep{2012MNRAS.423.1300R}. But we believe that more details need to be further considered,
such as a soft equation of state of NSs and the more practical case that the coupling of magnetic field and plasma in NSs.}.
However, if the central compact object is a black hole (BH), both processes may be correct.
So detecting associated GW radiation is helpful to understand the GRB physics.

\begin{figure}
\centering
  \includegraphics[width=0.65\textwidth]{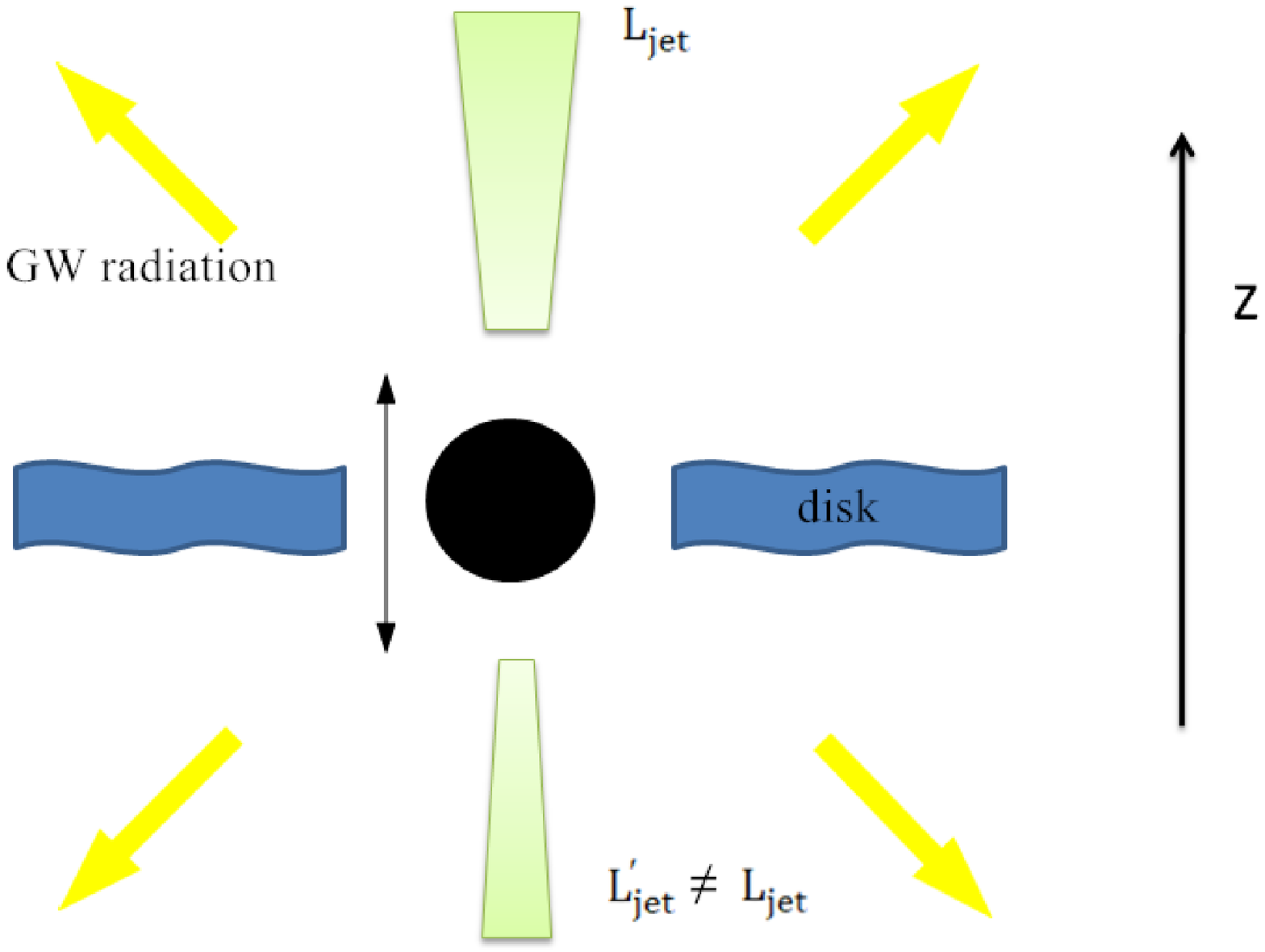}
  \includegraphics[width=0.65\textwidth]{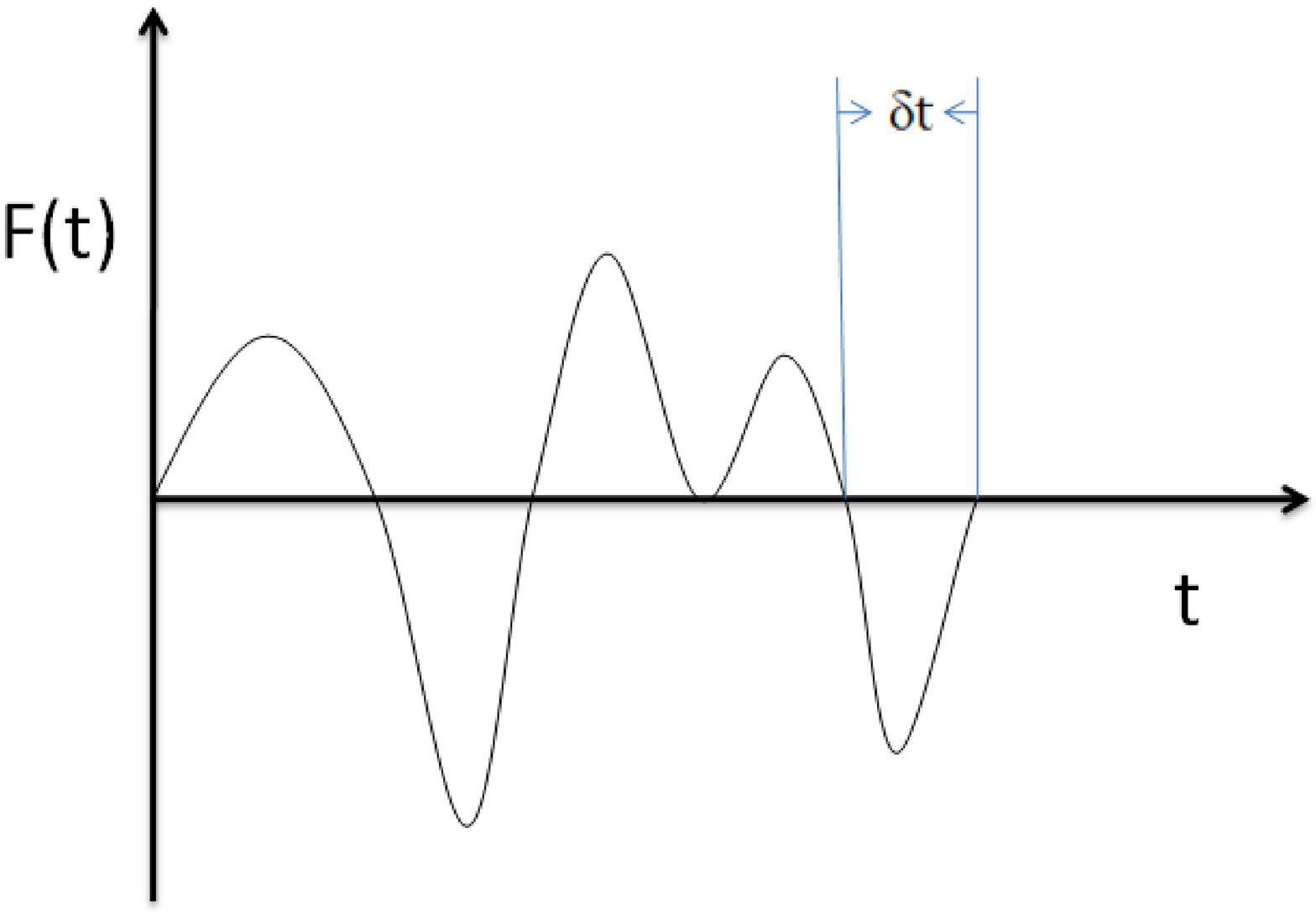}
  \caption{Upper panel: A schematic of GW production by GRBs with asymmetric jets. The asymmetric relativistic jets are represented by trapezoids. The central compact object,
  represented by filled black circle, then exhibits motion or oscillation (fine bidirectional arrow) due to reaction force.
  This produces gravitational wave radiation is marked by coarse solid.
  The wavy strips represent the accretion disk.
  Lower panel: A schematic plot of the time variation of $F(t)$, the force acted on the central object.
  The plot is a bit arbitrary because we don't have any detailed knowledge about the form of $F(t)$.
  The typical ``period'' of $F(t)$ is chosen to be of the same magnitude of ${\delta t}$, the minimum-variation time scale of the GRB light curves.\label{fig.1}}
\end{figure}

In this paper, we studied the GW production induced by the asymmetric jets of GRBs, and discuss its physical implications.
Since the collapses can be nonspherical in SNe, one can also imagine that the bipolar jets are asymmetric in GRBs (see the upper panel of Fig.\ref{fig.1}).
There are more clues suggesting the existence of asymmetry. As a thermodynamic system, the star-disk system inevitable suffers from thermodynamical fluctuation;
so in the process of accretion, the appearance of small asymmetry is reasonable. Also, as we can see from observations \citep{1995ARA&A..33..415F}
the light curves of GRBs are varied and irregular, implying the activities of the GRB central engines are inhomogeneous. Finally, in any case, it is worthy searching for GWs from GRBs to determine whether the asymmetry exists from the observational side. That will provide useful information for theoretical \citep{1999ApJ...518..356P,2006ApJ...643L..87G,2009ApJ...700.1970L} and simulation investigations \citep{2011MNRAS.410.2302Z, 2016ApJ...824L...6R,2017PhRvD..96d3006S}.

The asymmetric bipolar jets will have net recoil force on the central compact objects, and then induce the GW radiation.
This provides a new mechanism for the GW production in GRBs,
and is different from the scenario of accelerating relativistic jets \citep{2001PhRvD..64f4018S,2001astro.ph..2315P,2002grg..conf..259P,2004PhRvD..70j4012S}.

This paper is organized as follows.
In Sec. \ref{sec2} we study the properties of the produced GW, divided by the cases of NS and BH systems, respectively.
In Sec. \ref{sec3} we discuss possibility of detection by future experiments.
Sec. \ref{sec4} discusses the implication for GRB physics.
We conclude in Sec. \ref{sec5}.

\section{GW PRODUCTION IN GRBs WITH ASYMMETRIC JETs} \label{sec2}
Considering a compact GRB event emitting asymmetric jets.
The direct measurable quantity is the jet luminosity towards the earth $L_{\rm jet}(t)$.
Denoting the luminosity of the jet in the opposite direction as $L^{'}_{\rm jet}(t)$,
the asymmetry can be expressed as
\begin{eqnarray}
\Delta L_{\rm jet}(t)=L_{\rm jet}(t)-L^{'}_{\rm jet}(t)=\eta (t)L_{\rm jet}(t),
\end{eqnarray}
where we use $\eta (t)$ to characterize the magnitude of asymmetry.
Correspondingly, there will be a recoil force acted on the central compact object, i.e.
\begin{eqnarray}\label{Eq2}
F(t)\approx \eta (t)L_{\rm jet}(t)/c,
\end{eqnarray}
where $c$ is the speed of light. The motion of the star leads to emission of GW. A schematic illustration is shown in the upper panel of Fig.\ref{fig.1}.

In what follows we calculate the characteristic strain of the GW emitted by the system.
We take a Cartesian coordinate system of $x_i=(x_1,x_2,x_3)=(x,y,z)$, where the center of mass of the central compact objetct is taken as the origin,
and we choose the direction of the $z$-axis to be parallel to the direction of $F(t)$, as well as the direction of the jet motion.
We divide our discussion into cases of NS-disk and BH-disk systems, respectively.

For a NS-disk system, by adopting a weak gravitational field and Newtonian limit, the waveform under trace-reversed metric perturbation is \citep{2011gwpa.book.....C}
\begin{eqnarray}\label{Eq3}
\bar{h}_{i3}(t)\approx \frac{2G}{c^{4}D}\frac{\partial ^{2}}{{\partial t^{2}}}\int\rho x_{i}x_{3}d{V}.
\end{eqnarray}
Here $D$ is the distance from the wave source to the earth,
while $V$ and $\rho$ are the volume and mass density of the star, respectively.
The time evolution of $x_{3}$ satisfies
\begin{eqnarray}\label{Eq4}
\frac{d^{2}x_{3}}{dt^{2}}=\frac{F(t)}{M}=\frac{\eta (t)L_{\rm jet}(t)}{Mc},
\end{eqnarray}
where
\begin{eqnarray}\label{Eq5}
M=M_{0}+\int\dot{m}dt.
\end{eqnarray}
Here $M_{0}$ is the initial mass of the star, and $\dot{m}$ is mass accretion rate which can be estimated following the method of Du et al (2017).

For a BH-disk system, the weak gravitational field and Newtonian limit is invalid. To be strict, one should conduct the calculation fully relativistically (see Appendix \ref{app}).
In this work, since we only intend to give a rough estimation of the GW, we simply regard the BH as a bulk and treated dynamically just like the NS-disk case.
In addition to Eqs.(\ref{Eq3}, \ref{Eq4}, \ref{Eq5}), we have the following equations
\begin{eqnarray}
V=\frac{4}{3}\pi r_{+}^{3},
\end{eqnarray}
\begin{eqnarray}
r_{+}=\frac{GM}{c^{2}}(1+\sqrt{1-a_{\ast}^{2}}),
\end{eqnarray}
where $r_{+}$ and $a_{\ast}$ are the radius of the outer horizon and dimensionless angular momentum of the BH \citep{1998bhad.conf.....K}, respectively,
and the evolution of $a_{\ast}$ during accretion satisfies \citep{2017ApJ...849...47L}
\begin{eqnarray}
\frac{da_{\ast}}{dt}=\frac{\dot{m}L_{\rm ms}c}{GM^{2}}-\frac{2a_{\ast}\dot{m}E_{\rm ms}}{mc^{2}},
\end{eqnarray}
where $E_{\rm ms}=(4\sqrt{R_{\rm ms}-3a_{\ast}})/(\sqrt{3}R_{\rm ms})$, $L_{\rm ms}=(GM/c)(6\sqrt{R_{\rm ms}}-4a_{\ast})(\sqrt{3R_{\rm ms}})$,
and $R_{\rm ms}$ is the dimensionless radius of the marginally stable orbit of the BH (here the subscript ``ms'' stands for marginally stable orbit).
Usually, in this case, the mass of accretion disk is much smaller than the original mass $M_{0}$, so $M\approx M_{0}$.

\section{Possibility of detection by future experiments}\label{sec3}
\subsection{Estimation of Amplitude and frequency}\label{sec3.1}

Since we treat the BH-disk system as the NS-disk system,
simulating Eqs.(\ref{Eq2}, \ref{Eq3}, \ref{Eq4}), the amplitude of GWs for both cases can be estimate as
\begin{eqnarray}\label{Eq9}
\bar{h}&\sim& \frac{GR}{c^{4}D}\left ( \frac{\eta (t)L_{\rm jet}(t)}{c} \right )\nonumber \\
&=&1.8\times 10^{-26}\left ( \frac{\eta}{0.1} \right )\left ( \frac{R}{10\rm km} \right )\left ( \frac{D}{10\rm kpc} \right )^{-1}\left ( \frac{L_{\rm jet}}{10^{51}\rm erg\cdot s^{-1}} \right ),
\end{eqnarray}
where $R$ is the radius of the star coming from the integral over $x_i$ (for BHs, we have $R=r_{+}$).

To be intuitively clear, we expressed $h$ by characteristic values of $\eta$, $R$, $D$ and $L_{\rm jet}$.
We adopt $\eta(t)=0.1$ since this value will not affect the stability of the system (see Eq.\ref{Eq11}).
Under these values, the characteristic value of $\bar h$ is found to be $1.8 \times 10^{-26}$.

According to GRB light curves, the minimum time scale that the variation takes place is $\delta t\sim 1\rm ms$ \citep{2002MNRAS.330..920N}.
Considering the correlation between the variational light curves and the activity of the central engines,
one can reasonably assume that the reversal time scale of the recoil force $\Delta t$ being the same order of
magnitude of $\delta t$ (see the lower panel of Fig.\ref{fig.1}).
Besides, taking note of the duration of GRBs $T_{90}\in(0.1-100)\rm s$ and the stability of the systems,
the upper limit of the reversal time scale should be order of $\sim 1\rm s$
(see the discussion of section \ref{sec3.4}).
Then we estimate the frequency of GWs is
\begin{eqnarray}
f\sim (1-1000)\rm Hz.
\end{eqnarray}

A clear result is that, the characteristic $\bar{h}$ is too small to be detected by current observations.
The sensitivity of aLIGO is $\sim 10^{-23}$ in $(40-1000)\rm Hz$, while
the sensitivity of the third generation GW detector Einstein Telescope is
$\sim 10^{-24}$ in $(20-1000)\rm Hz$ \citep{2010CQGra..27s4002P}.
Both of them are far from the precision needed to detect our signal.

In what follows we discuss two possible mechanism enhancing the amplitude of $\bar h$ and makes it easier to be detected.

\subsection{Possible enhancement from Blandford-Znajek mechanism}\label{sec3.2}

For a BH-disk system, the relativistic jets may be launched by BZ mechanism.
The total power of the BZ process can be estimated as \citep{2000PhR...325...83L}
\begin{eqnarray}
L_{\rm BZ}=1.7\times 10^{50}a_{\ast }^{2}\left ( \frac{M}{1\rm M_{\odot }} \right )^{2}\left (\frac{B}{10^{15}\rm Gs} \right )^{2}F(a_{\ast})\rm erg\cdot s^{-1},
\end{eqnarray}
where
\begin{eqnarray}
F(a_{\ast})=\left ( \frac{1+\phi  ^{2}}{\phi^{2}} \right )\left [ \left ( \phi+\frac{1}{\phi} \right )\arctan\phi-1  \right ],
\end{eqnarray}
\begin{eqnarray}
\phi=\frac{a_{\ast}}{1+\sqrt{1-a_{\ast}^{2}}}.
\end{eqnarray}
$B$ is the strength of magnetic field of accretion disk, which is also a major uncertainty in estimating $L_{\rm BZ}$.
To estimate BZ power, a reasonable assumption is the magnetic pressure balances the ram pressure of the accretion flow,
then the BZ power can be expressed as \footnote{Bing Zhang, \emph{The Physics of Gamma-Ray Bursts}, Cambridge University Press, in press}
\begin{eqnarray}
L_{\rm BZ}=9.3\times 10^{53}a_{\ast}^{2}\left ( \frac{\dot{m}}{1\rm M_{\odot }s^{-1}} \right )X(a_{\ast})\rm erg\cdot s^{-1},
\end{eqnarray}
where
\begin{eqnarray}
X(a_{\ast})=F(a_{\ast})/\left(1+\sqrt{1-a_{\ast}^{2}}\right)^{2}.
\end{eqnarray}
For a hyper-accreting extreme Kerr hole (for example, $a_{\ast}=0.9$ and $\dot{m}=3\rm M_{\odot}s^{-1}$),
$L_{\rm jet}$ may be as high as $10^{54}\rm erg\cdot s^{-1}$, such that one has
\begin{eqnarray}\label{o1}
\bar{h}\sim 10^{-23}
\end{eqnarray}
by adopting the same typical values as Eq.\ref{Eq9}.

\subsection{Possible enhancement from neutrino emission}\label{sec3.3}
Under the NS-disk case, the accretion matter will impact on the surface of the NS.
The gravitational potential energy of accretion matter will be transformed into neutrino radiation energy in large part.
That's to say, the asymmetry is caused by neutrino emission not just the neutrino annihilation.
The luminosity of the former $L_{\rm \nu}$ can be about $10^{3}$ times larger than the latter $L_{\rm \bar{\nu}\nu}$ \citep{2010ApJ...718..841Z},
where $L_{\rm \bar{\nu}\nu}\approx L_{\rm jet}$.
Then one also has
\begin{eqnarray}\label{o2}
\bar{h}\sim 10^{-23}
\end{eqnarray} by adopting the same typical values as before.

There are also works \citep{2016ApJ...816L..30J,2017JPhG...44h4007P} from the simulation side suggesting that
the neutrino pair annihilation process may not be able to explain the central engine of short GRBs,
so there are still controversies. Here we do not try to determine which mechanism is more possible;
rather we propose a possibility to study the GRB physics by observing GWs from asymmetric GRBs.
\subsection{Stability of the system}\label{sec3.4}

The recoil force will make the central object deviate from the equilibrium position.
The displacement relative to the equilibrium position is
\begin{eqnarray}\label{Eq11}
\frac{1}{2}\frac{\eta(t) L_{\rm jet}(t) }{Mc}(\Delta t)^{2}\sim 1.8\times 10^{-2}{\rm m}\left ( \frac{\eta }{0.1} \right )
\left ( \frac{L_{\rm jet}}{10^{51}\rm erg\cdot s^{-1}} \right )\left ( \frac{\Delta t}{1\rm ms} \right )^{2}\left ( \frac{M}{1.4\rm M_{\odot }} \right )^{-1}.
\end{eqnarray}
Steady accretion requires the central object to slightly oscillate near its equilibrium position.
A simple estimation is that the displacement can not exceed the characteristic scale of the system
\footnote{A more reasonable estimation is demand the perturbation of potential energy on the accretion disk
to be smaller than the difference of the potential energy between the surface of the central object and the marginally stable orbit.},
i.e. $R$.
So the upper limit of $\Delta t$ is order of $1\rm s$ under $\eta =0.1$.

It is worth noting that, by adopting the typical values of Eq.\ref{Eq11},
the displacement seems so small that it will hardly have an impact on the potential energy of the accretion disk.
But this oscillation may lead to a dominant periodic (or quasi-period) mode in the GWs.
Because, 1) there is always a minimum time scale that the variation of light curves take place corresponding to the activities of accretion disks;
2) stochastic low frequency oscillation may cause the central object to deviate from its equilibrium position too much, so that the stability of the system will be destroyed.

\subsection{Signal-to-noise ratio}
\begin{figure}[htbp]
\centering
\includegraphics[width=0.8\textwidth]{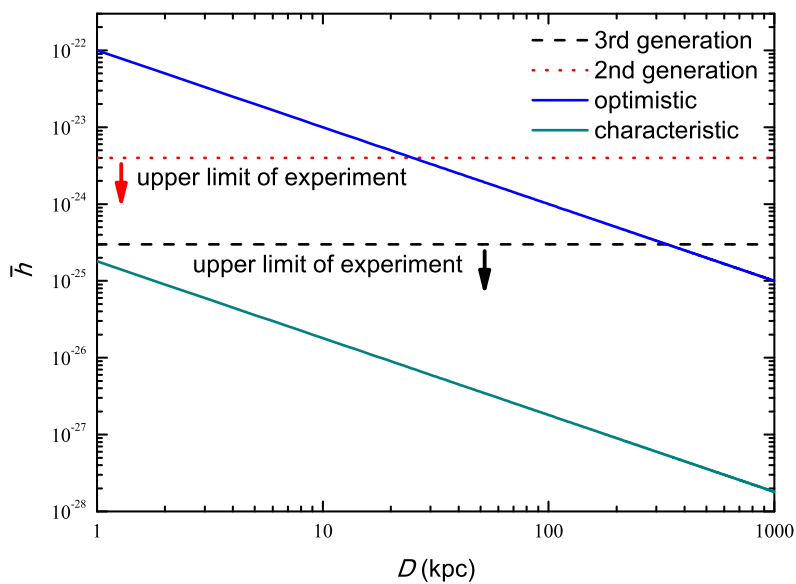}
\caption{The variation of the GW amplitude with distance. The dot line and dash line represent the upper limits of the sensitivity of aLIGO and Einstein Telescope \citep{2010CQGra..27s4002P}, respectively.
The optimistic estimation of GW (Eq.\ref{o1}, Eq.\ref{o2}) amplitude is shown as the blue solid line, while the characteristic GW amplitude is represented by the gray solid line.
One can see that the GWs induced by the asymmetric jets of GRBs may be detected by the third generation of GW detector. \label{fig.2}}
\end{figure}
It is beneficial to estimate the signal to noise ratio (SNR) of the GW, which is
\begin{eqnarray}
SNR^{2}=4\int_{0}^{\infty } \frac{\left | \widetilde{h}(f) \right |^{2}}{S_{h}(f)}df,
\end{eqnarray}
where $\tilde{h}(f)$ is the Fourier transformation of $\bar{h}(t)$, and $S_{h}(f)$ is the noise power spectral density of detectors.
According to section \ref{sec3.4}, if the GW signal satisfies $\bar{h}(t)=h_{0}\cos \omega t$, then we have
\begin{eqnarray}
SNR\sim \sqrt{T_{90}}\left ( \frac{h_{0}}{10^{-23}} \right )\left ( \frac{S_{h}\left ( \omega /2\pi  \right )}{10^{-46}\rm Hz^{-1}} \right )^{-1/2}.
\end{eqnarray}
The gravitational wave signal is detectable if the SNR exceeds the rough threshold of $5$.

As we see, the characteristic $\bar{h}$ is too small to be detected, but possibility still exists that it can be detected by future experiments, such as Einstein Telescope (see Fig.\ref{fig.2}).

\section{Implication for GRB physics}\label{sec4}
\subsection{Reconstructing activity of the central engine}\label{sec4.1}
One may worry that, without knowing the specific form of $F(t)$, it is impossible to discuss the quantitative results of these equations.
On the contrary, if these GWs are detected, the initial mass and angular momentum of the star can be inferred \citep{2013PhRvD..88f2001A}.
One can reconstruct the form of $F(t)$ according to Eqs. (4-9) (or the QNMs) under appropriate initial conditions, for example, $z(t=0)=z_{0}$ and ${dz(t=0)}/{dt}=0$.
The restructuring $F(t)$, which records the activity of the central engine, can be used to study GRB source physics.

For example, under the internal shock scenario \citep{1994ApJ...430L..93R}, the GRB central engines should be intermittently active.
If this model is true, the induced GWs will be also intermittent, as well as the recoil force. By reconstructing $F(t)$, this model can be test.

\subsection{Determining jets physics}
BH-disk systems may launch the relativistic jets through the BZ mechanism or through the neutrino annihilation process,
but only the former mechanism leads to GW production. The jets from neutrino annihilation process of BH-disk systems do not lead to GW production.
Unlike NS-disk systems, accretion matter falls directly into the BH, and does not stay on its ``surface'',
such that the BH does not radiate neutrinos due to the collision between BH surface and accretion matter.
Therefore, if the neutrino annihilation process is ture, the jet can only originate from the neutrino annihilation process on the accretion disk, and the central BH will not have GW radiation.
On the other hand, the BZ mechanism directly extract the rotation energy of the BH through its ergosphere, has a recoil force to the BH.
At this time, there should be a GW radiation. So, by detecting the GWs induced by the asymmetry jets, one can screen out the launch mechanism of these jets.

\section{CONLUSION}\label{sec5}

We studied the properties of GWs production from asymmetric jets in GRBs, and found that these GWs
have the memory of the activities of GRB central engines.
The phenomenon discussed by this paper is expected to exist, as long as no mechanism ensures that the jets of
GRBs being symmetric. A lot of information about the system can be inferred via the property of the produced GWs.
Especially, detection of these GWs may be the only way for us to determine the long-standing problems about the GRB relativistic jets,
as well as for other basic physical problems \citep{2016ApJ...827...75L}.

From the viewpoint of practical application, there are two limitations (similar to the case of SNe).
The first is the amplitude of these GWs is below the sensitivity of current GW detectors.
Even with best expectations, these GWs can only be detected by the next generation of detectors.
The second is the event rate of GRBs is quite low.
Even if all GRBs are associated with core-collapse SNe and kilonovae,
the detection rate of the associated GW event is only about $(2\sim3) \cdot\rm (100kpc)^{-3}(100yr)^{-1}$
\citep{2010MNRAS.409L.132Z,2016PhRvD..94j2001A,2017ApJ...834...84C}.
More developments in the GW experiments are demanded to study\citep{2018MNRAS.tmp.1001C}.

Although the signal is difficult to be detected, considering the significant implications to GRB physics,
it is always worthy searching the emitted GWs in experiments.
We hope this goal can be achieved in not very far future.

\section{Acknowledgement}
We would like to thank the anonymous referee for
his/her very useful comments that have allowed us to
improve our paper.
We thank Yi-Jung Yang for useful discussion.
This work is supported by the National Natural
Science Foundation of China (Grant No. 11703098, Grant No. 11275247, and
Grant No. 11335012) and a 985 grant at Sun Yat-Sen University.
Y.-M. Hu gratefully acknowledge the support of the National Natural Science Foundation of China (1170030038).
F. K. Peng acknowledges support from the Doctoral Starting up Foundation of Guizhou Normal University 2017 (GZNUD[2017] 33).

\begin{appendix}
\section{Appendix} \label{app}
The mass of accretion disk is usually much smaller than the central BH.
The effect of the small accretion mass on the BH can be regarded as a perturbation.
This perturbation can be described by the Teukolsky equation, whose solutions are known as quasi-normal modes (QNMs) \citep{1972PhRvL..29.1114T,1973ApJ...185..635T}.

As discussed in Section \ref{sec3.4}, the oscillation of the BH may be quasi periodic.
If there is a sinusoidal oscillation, this indicates that the leading mode of QNMs is the fundamental $l=2$ quasi-normal mode.

Following the method of Creighton $\&$ Anderson (2011), the amplitude can be express as
\begin{eqnarray}\label{A1}
\bar{h}\sim 10^{-23} \left ( \frac{\varepsilon }{10^{-6}} \right )\left (  \frac{D}{10\rm kpc}\right )^{-1}\left ( \frac{\dot{m}}{1\rm M_{\odot }s^{-1}} \right ).
\end{eqnarray}
Since it is difficult to estimate the GW radiation efficiency $\varepsilon$, we find it hard to estimate the GW amplitude.

\end{appendix}


\begin{thebibliography}{}

\bibitem[\protect\citeauthoryear{Aasi et al.}{2013}]{2013PhRvD..88f2001A} Aasi J., et al., 2013, PhRvD, 88, 062001


\bibitem[\protect\citeauthoryear{Abbott et al.}{2017a}]{2017ApJ...848L..13A} Abbott B.~P., et al., 2017a, ApJ, 848, L13


\bibitem[\protect\citeauthoryear{Abbott et al.}{2017b}]{2017ApJ...851L..16A} Abbott B.~P., et al., 2017b, ApJ, 851, L16


\bibitem[\protect\citeauthoryear{Abbott et al.}{2016}]{2016PhRvD..94j2001A} Abbott B.~P., et al., 2016, PhRvD, 94, 102001


\bibitem[\protect\citeauthoryear{Arcavi et al.}{2017}]{2017Natur.551...64A} Arcavi I., et al., 2017, Natur, 551, 64


\bibitem[\protect\citeauthoryear{Blandford \& Znajek}{1977}]{1977MNRAS.179..433B} Blandford R.~D., Znajek R.~L., 1977, MNRAS, 179, 433


\bibitem[\protect\citeauthoryear{Chan et al.}{2017}]{2017ApJ...834...84C} Chan M.~L., Hu Y.-M., Messenger C., Hendry M., Heng I.~S., 2017, ApJ, 834, 84


\bibitem[\protect\citeauthoryear{Coughlin et al.}{2018}]{2018MNRAS.tmp.1001C} Coughlin M.~W., et al., 2018, MNRAS,


\bibitem[\protect\citeauthoryear{Coulter et al.}{2017}]{2017Sci...358.1556C} Coulter D.~A., et al., 2017, Sci, 358, 1556


\bibitem[\protect\citeauthoryear{Creighton \& Anderson}{2011}]{2011gwpa.book.....C} Creighton J., Anderson W., 2011, gwpa.book,


\bibitem[\protect\citeauthoryear{Du et al.}{2017}]{2017arXiv171205964D} Du S., Peng F.-K., Long G.-B., Li M., 2017, arXiv, arXiv:1712.05964


\bibitem[\protect\citeauthoryear{Fishman \& Meegan}{1995}]{1995ARA&A..33..415F} Fishman G.~J., Meegan C.~A., 1995, ARA\&A, 33, 415


\bibitem[\protect\citeauthoryear{Gu, Liu, \& Lu}{2006}]{2006ApJ...643L..87G} Gu W.-M., Liu T., Lu J.-F., 2006, ApJ, 643, L87


\bibitem[\protect\citeauthoryear{Hayama et al.}{2016}]{2016PhRvL.116o1102H} Hayama K., Kuroda T., Nakamura K., Yamada S., 2016, PhRvL, 116, 151102


\bibitem[\protect\citeauthoryear{Just et al.}{2016}]{2016ApJ...816L..30J} Just O., Obergaulinger M., Janka H.-T., Bauswein A., Schwarz N., 2016, ApJ, 816, L30


\bibitem[\protect\citeauthoryear{Kato, Fukue, \& Mineshige}{1998}]{1998bhad.conf.....K} Kato S., Fukue J., Mineshige S., 1998, bhad.conf,


\bibitem[\protect\citeauthoryear{Kumar \& Zhang}{2015}]{2015PhR...561....1K} Kumar P., Zhang B., 2015, PhR, 561, 1


\bibitem[\protect\citeauthoryear{Lee, Wijers, \& Brown}{2000}]{2000PhR...325...83L} Lee H.~K., Wijers R.~A.~M.~J., Brown G.~E., 2000, PhR, 325, 83


\bibitem[\protect\citeauthoryear{Lei et al.}{2017}]{2017ApJ...849...47L} Lei W.-H., Zhang B., Wu X.-F., Liang E.-W., 2017, ApJ, 849, 47


\bibitem[\protect\citeauthoryear{Lei et al.}{2009}]{2009ApJ...700.1970L} Lei W.~H., Wang D.~X., Zhang L., Gan Z.~M., Zou Y.~C., Xie Y., 2009, ApJ, 700, 1970


\bibitem[\protect\citeauthoryear{Li et al.}{2016}]{2016ApJ...827...75L} Li X., Hu Y.-M., Fan Y.-Z., Wei D.-M., 2016, ApJ, 827, 75


\bibitem[\protect\citeauthoryear{Liu, Gu, \& Zhang}{2017}]{2017NewAR..79....1L} Liu T., Gu W.-M., Zhang B., 2017, NewAR, 79, 1


\bibitem[\protect\citeauthoryear{MacFadyen \& Woosley}{1999}]{1999ApJ...524..262M} MacFadyen A.~I., Woosley S.~E., 1999, ApJ, 524, 262


\bibitem[\protect\citeauthoryear{Nakar \& Piran}{2002}]{2002MNRAS.330..920N} Nakar E., Piran T., 2002, MNRAS, 330, 920


\bibitem[\protect\citeauthoryear{Perego, Yasin, \& Arcones}{2017}]{2017JPhG...44h4007P} Perego A., Yasin H., Arcones A., 2017, JPhG, 44, 084007


\bibitem[\protect\citeauthoryear{Popham, Woosley, \& Fryer}{1999}]{1999ApJ...518..356P} Popham R., Woosley S.~E., Fryer C., 1999, ApJ, 518, 356


\bibitem[\protect\citeauthoryear{Piran}{2001}]{2001astro.ph..2315P} Piran T., 2001, astro, arXiv:astro-ph/0102315


\bibitem[\protect\citeauthoryear{Piran}{2002}]{2002grg..conf..259P} Piran T., 2002, grg..conf, 259


\bibitem[\protect\citeauthoryear{Punturo et al.}{2010}]{2010CQGra..27s4002P} Punturo M., et al., 2010, CQGra, 27, 194002


\bibitem[\protect\citeauthoryear{Rees \& Meszaros}{1994}]{1994ApJ...430L..93R} Rees M.~J., Meszaros P., 1994, ApJ, 430, L93


\bibitem[\protect\citeauthoryear{Ruiz et al.}{2012}]{2012MNRAS.423.1300R} Ruiz M., Palenzuela C., Galeazzi F., Bona C., 2012, MNRAS, 423, 1300


\bibitem[\protect\citeauthoryear{Ruiz et al.}{2016}]{2016ApJ...824L...6R} Ruiz M., Lang R.~N., Paschalidis V., Shapiro S.~L., 2016, ApJ, 824, L6


\bibitem[\protect\citeauthoryear{Sago et al.}{2004}]{2004PhRvD..70j4012S} Sago N., Ioka K., Nakamura T., Yamazaki R., 2004, PhRvD, 70, 104012


\bibitem[\protect\citeauthoryear{Segalis \& Ori}{2001}]{2001PhRvD..64f4018S} Segalis E.~B., Ori A., 2001, PhRvD, 64, 064018


\bibitem[\protect\citeauthoryear{Sun et al.}{2017}]{2017PhRvD..96d3006S} Sun L., Paschalidis V., Ruiz M., Shapiro S.~L., 2017, PhRvD, 96, 043006


\bibitem[\protect\citeauthoryear{Teukolsky}{1973}]{1973ApJ...185..635T} Teukolsky S.~A., 1973, ApJ, 185, 635


\bibitem[\protect\citeauthoryear{Teukolsky}{1972}]{1972PhRvL..29.1114T} Teukolsky S.~A., 1972, PhRvL, 29, 1114


\bibitem[\protect\citeauthoryear{Zalamea \& Beloborodov}{2011}]{2011MNRAS.410.2302Z} Zalamea I., Beloborodov A.~M., 2011, MNRAS, 410, 2302


\bibitem[\protect\citeauthoryear{Zhang \& Dai}{2010}]{2010ApJ...718..841Z} Zhang D., Dai Z.~G., 2010, ApJ, 718, 841


\bibitem[\protect\citeauthoryear{Zhang \& Yan}{2011}]{2011ApJ...726...90Z} Zhang B., Yan H., 2011, ApJ, 726, 90


\bibitem[\protect\citeauthoryear{Zhu, Howell, \& Blair}{2010}]{2010MNRAS.409L.132Z} Zhu X.-J., Howell E., Blair D., 2010, MNRAS, 409, L132




\end{thebibliography}
\end{document}